\documentclass[prd,superscriptaddress,twocolumn,nofootinbib,showpacs,preprintnumbers,floatfix]{revtex4}

\usepackage{mathrsfs}
\usepackage{amsfonts,amssymb,amsmath}
\usepackage{graphicx,epsfig}
\usepackage{multirow}

\def\fsu5{$\cal{F}$-$SU(5)$}
\def\bfsu5{$\boldsymbol{\mathcal{F}}$-$\boldsymbol{SU(5)}$}

\def\m1half{$M_{1/2}$}
\def\m3half{$M_{3/2}$}
\def\m32{$M_{32}$}

\def\fb{${\rm fb}^{-1}$~}

\def\mt2{$M_{T2}$}
\def\x2{$\chi^2$}
\def\2b{$M_{T2}b$}

\def\bs0{$B_S^0 \rightarrow \mu^+ \mu^-$}

\begin{document}

\title{The Race for Supersymmetric Dark Matter at XENON100 and the LHC:
\\Stringy Correlations from No-Scale $\cal{F}$-$SU(5)$}

\author{Tianjun Li}

\affiliation{State Key Laboratory of Theoretical Physics, Institute of Theoretical Physics,
Chinese Academy of Sciences, Beijing 100190, P. R. China }

\affiliation{George P. and Cynthia W. Mitchell Institute for Fundamental Physics and Astronomy,
Texas A$\&$M University, College Station, TX 77843, USA }

\author{James A. Maxin}

\affiliation{George P. and Cynthia W. Mitchell Institute for Fundamental Physics and Astronomy,
Texas A$\&$M University, College Station, TX 77843, USA }

\author{Dimitri V. Nanopoulos}

\affiliation{George P. and Cynthia W. Mitchell Institute for Fundamental Physics and Astronomy,
Texas A$\&$M University, College Station, TX 77843, USA }

\affiliation{Astroparticle Physics Group, Houston Advanced Research Center (HARC),
Mitchell Campus, Woodlands, TX 77381, USA}

\affiliation{Academy of Athens, Division of Natural Sciences,
28 Panepistimiou Avenue, Athens 10679, Greece }

\author{Joel W. Walker}

\affiliation{Department of Physics, Sam Houston State University,
Huntsville, TX 77341, USA }


\begin{abstract}

The discovery of supersymmetry (SUSY) via action of the cold dark matter
candidate is being led on the indirect collider production front by the LHC, and correspondingly
by the XENON100 collaboration on the direct detection front.  We undertake a dual case study of
the specific SUSY signatures which the No-Scale flipped $SU(5)\times U(1)_X$ grand unified theory
with TeV-scale vector-like particles (No-Scale $\cal{F}$-$SU(5)$) would exhibit at each of these 
experiments.  We demonstrate a correlation between the near-term
prospects of these two distinct approaches.  We feature a dark matter candidate which
is over 99\% bino due to a comparatively large Higgs bilinear mass $\mu$ term around the electroweak scale, and thus
automatically satisfy the current constraints from the XENON100 and CDMS/EDELWEISS experiments.
We do however expect that the ongoing extension of the XENON100 run may effectively probe our model.
Likewise, our model is also currently being probed by the LHC via a search for events with ultra-high multiplicity hadronic jets,
which are a characteristic feature of the distinctive No-Scale $\cal{F}$-$SU(5)$ mass hierarchy.

\end{abstract}


\preprint{ACT-07-11, MIFPA-11-20}

\maketitle


\section{Introduction}

The possibility that common baryonic matter represents only a small fraction of our Universe's net positive
non-gravitational energy budget was first suggested in 1933 by Zwicky~\cite{Zwicky:1933gu}, motivated by anomalies in
the comparison of observed galactic rotation curves to predictions based on visible stellar content.  The notion of
dark matter (DM) has since become a pillar of modern cosmology, lent independent support by detailed measurement
of $\mathcal{O}\,(10^{-5})$ anisotropies in the Cosmic Microwave Background (CMB).  Indeed, the proportional allocation by
the WMAP~\cite{Spergel:2003cb,Spergel:2006hy,Komatsu:2010fb} satellite of baryonic matter $\simeq 4\%$, cold dark matter (CDM) $\simeq 23\%$, and dark
energy (a cosmological constant) $\simeq 73\%$ within a cosmologically flat unit normalization is now axiomatically familiar.

The essential non-baryonic character of the vast majority of non-luminous matter is inferred i) by extrapolating from the
current abundance ratios of light isotopes back to the required density of baryonic particles during big bang nucleosynthesis,
and ii) by studying the residual imprint onto the CMB acoustic peaks of interactions prior to decoupling with the photonic
radiation of the primordial plasma.  The fact that dark matter should be primarily cold, {\it i.e.}~possessing a non-relativistic
thermal speed at the time of its own decoupling, is forced primarily by the requirements of large scale structure formation.
Highly relativistic matter cannot clump sufficiently to account for the ``bottom-up'', or small-to-big creation of
structure which observation favors.  Of course, the requirement that the 
cold DM (CDM) is indeed dark, {\it i.e.}~that it not scatter
electromagnetic radiation, further implies charge neutrality~\cite{Ellis:1983ew}.
A stable Weakly Interacting Massive Particle (WIMP), having a canonical weak-scale mass, may elegantly account for a
thermal CDM relic density of the correct magnitude order~\cite{Ellis:1983ew}.
Although the Standard Model (SM) contains no particle which may represent the CDM, extensions featuring Supersymmetry (SUSY) --
as independently essential for stabilization of the electroweak (EW) mass hierarchy against devastating Planck-scale quantum corrections --
do provide natural candidates in the form of the lightest supersymmetric particle (LSP)~\cite{Goldberg:1983nd}\cite{Ellis:1983ew},
protected from decay by conservation of $R$-parity.

The Large Hadron Collider (LHC) at CERN has been steadily accumulating data from ${\sqrt s}=7-8$ TeV proton-proton collisions
since March 2010 at the CMS and ATLAS Experiments, and is projected to reach an integrated luminosity of about $25~{\rm fb}^{-1}$ by the end of 2012. However, early LHC results have produced no definitive signal of supersymmetry, severely constraining the experimentally viable parameter space of the CMSSM and mSUGRA~\cite{Chamseddine:1982jx} (for example, see Refs.~\cite{Strumia:2011dv,Baer:2012uy,Buchmueller:2012mc}). The absence of a hint of supersymmetry to date has advanced the constraints on the viable CMSSM and mSUGRA model space, posing the question of whether there exist SUSY and/or superstring post-Standard Model extensions that can evade the LHC constraints thus far imposed, though remaining within the current and near-term reach of the LHC.

Meanwhile however, there is a credible dark horse in the chase for dark matter, that being the much smaller scale direct detection experiments. This effort is being led by the XENON100 collaboration~\cite{Aprile:2011hi,Aprile:2012nq}, 
whose expansion in the last year to a fiducial detector mass of 62 kg of ultra-pure
liquid xenon has promptly netted a near ten-fold improvement over 
the CDMS and EDELWEISS experiments~\cite{CDMS:2011gh} in the upper bound on the spin-independent
cross section for scattering WIMPs against nucleons.  This limit
likewise begins to cut incisively against the favored regions of the CMSSM.

The exploration of the existence of dark matter persists not only between parallel experimental search strategies, but also between
competing theoretical proposals.  Recently, we have studied in some substantial detail a promising model by the name of No-Scale 
$\cal{F}$-$SU(5)$~\cite{Li:2010ws, Li:2010mi,Li:2010uu,Li:2011dw, Li:2011hr, Maxin:2011hy, Li:2011xu,Li:2011gh,Li:2011rp,Li:2011fu,Li:2011xg,Li:2011ex,Li:2011av,Li:2011ab,Li:2012hm,Li:2012tr,Li:2012ix,Li:2012yd,Li:2012qv,Li:2012jf}, which 
is constructed from the merger of the ${\cal F}$-lipped $SU(5)$ Grand Unified Theory
(GUT)~\cite{Barr:1981qv,Derendinger:1983aj,Antoniadis:1987dx},
two pairs of hypothetical TeV scale vector-like supersymmetric multiplets with origins in
${\cal F}$-theory~\cite{Jiang:2006hf,Jiang:2009zza,Jiang:2009za,Li:2010dp,Li:2010rz},
and the dynamically established boundary conditions of No-Scale
Supergravity~\cite{Cremmer:1983bf,Ellis:1983sf, Ellis:1983ei, Ellis:1984bm, Lahanas:1986uc}.
The viable parameter space of No-Scale $\cal{F}$-$SU(5)$ has been comprehensively mapped~\cite{Li:2011xu}, which satisfies the
``bare minimal'' phenomenological constraints, gives the correct CDM relic density, and is consistent
with a dynamic determination by the secondary minimization of the Higgs potential via the ``Super No-Scale''
mechanism~\cite{Li:2010uu,Li:2011dw,Li:2011xu,Li:2011ex}.

In the present work we undertake as a dual case study the specific elaboration of what detection of a unified SUSY DM signal emerging
from this preferred region would look like at XENON100 and the LHC. An unambiguous declaration of the discovery of supersymmetric
dark matter will actually demand complementary breakthroughs from both experiments.  Specifically, XENON100 and the direct
detection searches can only intrinsically claim detection of a WIMP, and cannot alone confirm the identification as a SUSY neutralino.
Conversely, the LHC and collider experiments may potentially observe the standard SUSY signal of missing energy plus
hadronic jets only by letting the target DM candidate pairwise escape the detector.

Thus, we believe it is essential to assess the correlations tangibly present between both experimental approaches.  In order to comprehensively ascertain the interconnection between the direct and indirect detection methodologies, the supersymmetric model under study must have the capacity to support and allow the discovery of such a correlation.  We present here a model that naturally admits the essential
linkage necessary for delineating the direct and indirect detection interrelationship, and shall employ it to numerically
and graphically demonstrate a clearly defined and testable example of LHC and XENON100 complementarity.


\section{The Current State of Dark Matter Detection}

WIMPs have been searched for extensively, by collider experiments such
as LEP, as well as other experiments at the Tevatron and LHC, and by direct detection experiments such as CDMS~\cite{Ahmed:2009zw}, 
EDELWEISS~\cite{:2011cy}, LUX~\cite{Fiorucci:2009ak}, and XENON100~\cite{Aprile:2011hx}.  The former look for the characteristic
missing energy signal attributable to an escaping WIMP produced {\it in situ}, while the latter, shielded deep
underground, attempt to register the scintillation induced by recoil of heavy nuclei within a stationary target
from the scattering of environmental DM particles.

CDM candidates which mix strongly with the $Z$ boson have already been excluded.
For a WIMP mass of 90 GeV, the CDMS and EDELWEISS experiments have placed an upper bound on the
DM-nucleon cross section  of $3.3\times 10^{-44}~{\rm cm^2}$ ($3.3\times 10^{-8}~{\rm pb}$) at the 90\%
Confidence Level (CL)~\cite{CDMS:2011gh}.
In addition, the XENON100 experiment has probed the spin-independent
DM-nucleon cross section down to  $2\times 10^{-45}~{\rm cm^2}$ ($2\times 10^{-9}~{\rm pb}$) 
for a WIMP mass of 55 GeV~\cite{Aprile:2012nq}, also at the 90\% CL.  
Three candidate events were observed in the XENON100 study cited, but given the background
expectation of $(1.8\pm0.6)$ events, it does not constitute statistically significant
evidence of actual DM~\cite{Aprile:2011hi}.  Absent still a definitive signal, 
the CDMS, EDELWEISS, and XENON100 experiments have already given very strong constraints on some
viable parameter space of the supersymmetric standard model (SSM)~\cite{Akula:2011dd, Farina:2011bh}.
For example, the ``well-tempered'' LSP neutralino scenario~\cite{ArkaniHamed:2006mb} and 
the ``focus-point'' region~\cite{Chan:1997bi, Feng:1999mn, Feng:1999zg} are now highly disfavored,
since the higgsino components of the LSP neutralino are generically too large.

The high energies accessible to the LHC have significantly relaxed the kinematic restrictions on the available
phase space of SUSY production processes, and the enlarged collision cross-sections allowed the
ATLAS and CMS collaborations to rapidly
match and overtake long standing Tevatron and LEP limits.  Concern has been raised that such a significant reach into the parameter space of the CMSSM,
having thus far observed no signal of supersymmetry, bodes ill for the SUSY search at large.  This seems
still quite premature. Our perspective rather, is that at such a critical juncture in the search for dark matter, when two experiments of unprecedented scope each extend their reach into the model space where dark matter can plausibly be detected, the search for phenomenologically consistent and theoretically efficient models which can survive the advancing experimental constraints, while making clear and testable
predictions, takes on an elevated relevance and immediacy.

The exacting WMAP7~\cite{Komatsu:2010fb} measurements of the CDM relic density constitute a relatively strong constraint,
and the construction of a model which makes compliant predictions {\it naturally} is quite non-trivial.
In particular, the identity of the LSP field itself is critical to the size of the predicted annihilation
cross-sections.  In the SSM, there are several DM candidates, such as the LSP neutralino, gravitino, and sneutrino, etc.
where the LSP neutralino has been studied extensively. There are
four neutralinos ${\widetilde{\chi}_i^{0}}$ in the minimal SSM, which are 
the mass eigenstates for the half-spin superpartners
to the neutral $(B,W^0)$ gauge fields and scalar Higgs fields, 
{\it i.e.}~the bino $\widetilde{B}$, the wino $\widetilde{W}^0$ 
and the two
higgsinos $\widetilde{H}_u^0$ and $\widetilde{H}_d^0$~\cite{Ellis:1982xd,Ellis:1982xz,Ellis:1983wd}\cite{Ellis:1983ew}.  The physical LSP $\widetilde{\chi}^0_1$
will represent the lightest linear combination of these four fields~\cite{Ellis:1982xd,Ellis:1982xz,Ellis:1983wd}\cite{Ellis:1983ew}.
In general, the bino-component dominant LSP neutralino in isolation overshoots
the correct CDM relic density, having a characteristically small annihilation cross section.  By contrast, 
the wino-component or higgsino-component dominant LSP neutralino, which has a large annihilation cross section,
tends to make relic density predictions which are too small.  The selection of a suitable admixture may be considered an unwelcome source of fine tuning.

To compensate for this effect, a WIMP mass of a few TeV is generically required if the dominant
component of the LSP neutralino is wino or higgsino.  However, the SUSY breaking scale is then similarly enhanced,
meaning that yet another element of fine-tuning is required in order to obtain the correct EW scale. On the contrary,
if the LSP is dominantly bino flavored, then a suppression of the DM relic density is required,
and there are three well established solution classes available: i) the bulk region with a restriction
to moderate supersymmetry breaking soft masses~\cite{Goldberg:1983nd}\cite{Ellis:1983ew}, ii) the coannihilation region
which can extend upward to relatively large supersymmetry breaking soft masses~\cite{Ellis:1998kh,Griest:1990kh}, and iii)
the Higgs pole region~\cite{Drees:1992am}. The mechanism of coannihilation, representing the mutually associated annihilation of two SUSY particles,
is quite appealing, and is often invoked to dilute the residual DM. Nonetheless, this approach requires that some conspiracy of near mass-degeneracy occur between the neutral LSP and the next-to-the-lightest superpartner (NLSP). Although some suitable narrow strip of parameter space may often be delineated, within say the $(M_{1/2},M_0)$ plane of an mSUGRA based construction, this again could suggest fine tuning.

It is essential to recognize though that the onus of proof for the fine tuning charge is not satisfied solely by the {\it appearance}
of tuning in the final state numerics.  Even a model which relies upon an apparently specific spectral distribution
may be classified as natural {\it if} a system of dynamics, driven by well motivated symmetries, uniquely focuses predictions toward
the desirable outcome, or at least provides for an agreeably broad solution space.
One positive step toward satisfaction of this objective would be the establishment of a
dynamically driven predilection toward a bino dominated LSP.  In this example, the limits from CDMS,
EDELWEISS, and XENON100 are naturally relaxed since only the higgsinos couple directly to the $Z$ boson.
The question of whether there exist well motivated supersymmetric models that can satisfy the current CDM
direct detection experiments, yet are still potentially testable within the near-term future of the LHC and
XENON100 experiments, is clearly then one of great importance.


\section{No-Scale ${\cal F}$-$SU(5)$}


The No-Scale $\cal{F}$-$SU(5)$ construction inherits all of the most beneficial phenomenology~\cite{Nanopoulos:2002qk}
of flipped $SU(5)$~\cite{Barr:1981qv,Derendinger:1983aj,Antoniadis:1987dx}, as well as all of the valuable theoretical motivation of No-Scale
Supergravity~\cite{Cremmer:1983bf,Ellis:1983sf, Ellis:1983ei, Ellis:1984bm, Lahanas:1986uc},
including a deep connection to the string theory infrared limit
(via compactification of the weakly coupled heterotic theory~\cite{Witten:1985xb} or 
M-theory on $S^1/Z_2$ at the leading order~\cite{Li:1997sk}),
and a mechanism for SUSY breaking which preserves a vanishing cosmological constant at the tree level
(facilitating the observed longevity and cosmological flatness of our Universe~\cite{Cremmer:1983bf}).

The gauge group of flipped $SU(5)$ is $SU(5)\times U(1)_{X}$, which can be embedded into $SO(10)$.
The generator $U(1)_{Y'}$ is defined for fundamental five-plets as $+1/2$ for the doublet, and $-1/3$ for the triplet members.
The hypercharge is given by $Q_{Y}=( Q_{X}-Q_{Y'})/5$.  There are three families of Standard Model (SM) fermions,
whose quantum numbers under the $SU(5)\times U(1)_{X}$ gauge group are
\begin{equation}
F_i={\mathbf{(10, 1)}} \quad;\quad {\bar f}_i={\mathbf{(\bar 5, -3)}} \quad;\quad {\bar l}_i={\mathbf{(1, 5)}},
\label{eq:smfermions}
\end{equation}
where $i=1, 2, 3$.  For breaking the GUT symmetry, there is a pair of ten-plet Higgs, as well as a pair
of five-plet Higgs for electroweak symmetry breaking (EWSB).
\begin{eqnarray}
& H={\mathbf{(10, 1)}}\quad;\quad~{\overline{H}}={\mathbf{({\overline{10}}, -1)}} & \nonumber \\
& h={\mathbf{(5, -2)}}\quad;\quad~{\overline h}={\mathbf{({\bar {5}}, 2)}} &
\label{eq:Higgs}
\end{eqnarray}
Since there have been no observations of mass degenerate superpartners for the known SM fields, SUSY itself must be broken around the TeV scale.
In mSUGRA~\cite{Chamseddine:1982jx}, this begins in a hidden sector, and the secondary propagation by gravitational interactions into the observable sector is parameterized by universal SUSY-breaking ``soft terms'' that comprise the gaugino mass $M_{1/2}$, scalar mass
$M_0$ and the trilinear coupling $A$. The ratio of the low energy Higgs vacuum expectation values (VEVs) tan$\beta$, and the sign of
the SUSY-preserving Higgs bilinear mass term $\mu$ remain undetermined, while the magnitude of the $\mu$ term and its bilinear soft term $B_{\mu}$
are determined by the $Z$-boson mass $M_Z$ and tan$\beta$ after EWSB.  Considering only the most simple No-Scale scenario, $M_0$=A=$B_{\mu}$=0 at the boundary of unification, while the full array of low energy SUSY breaking soft-terms evolve down 
from the only non-zero parameter $M_{1/2}$. Consequently, the SUSY mass spectrum is proportional to $M_{1/2}$ at leading order,
which renders the bulk ``internal'' physical properties invariant under an overall rescaling.

The condition on consistency between the high-energy $B_\mu = 0$ and low-energy value of $B_\mu$ that is required by EWSB is very challenging to reconcile under the renormalization group equation (RGE) running.  Our solution involves modifications to the $\beta$-function coefficients that are generated by the inclusion of the extra vector-like flippon multiplets, that actively participate in radiative loops above their mass threshold $M_{\rm V}$. The mass $M_{\rm V}$ should be of the TeV order as suggested by naturalness in view of the gauge hierarchy and $\mu$ problems. Evading a Landau pole for the strong coupling constant constrains the set of vector-like flippon multiplets which may be
given a mass in this range to only two constructions with flipped charge assignments, which have been specifically realized
in the $F$-theory model building context~\cite{Jiang:2006hf,Jiang:2009zza, Jiang:2009za}.  We adopt the multiplets
\begin{equation}
{XF}_{\mathbf{(10,1)}} \equiv (XQ,XD^c,XN^c) \quad;\quad {\overline{Xl}}_{\mathbf{(1, 5)}} \equiv XE^c \,
\label{eq:flippons}
\end{equation}
where $XQ$, $XD^c$, $XE^c$ and $XN^c$ carry the same quantum numbers as the quark doublet, right-handed down-type quark,
charged lepton and neutrino, respectively.  Alternatively, the pair of $SU(5)$ singlets can be discarded, but phenomenological consistency then
necessitates the considerable implementation of unspecified GUT thresholds.  Either way, the (formerly negative) one-loop $\beta$-function
coefficient of the strong coupling $\alpha_3$ becomes precisely zero, flattening the RGE running, and generating a wide
gap between the large $\alpha_{32} \simeq \alpha_3(M_{\rm Z}) \simeq 0.11$ and the much smaller $\alpha_{\rm X}$ at the scale $M_{32}$ of the intermediate flipped $SU(5)$ unification of the $SU(3)_C \times SU(2)_{\rm L}$ subgroup, facilitating a quite important secondary running phase
up to the final $SU(5) \times U(1)_{\rm X}$ unification scale~\cite{Li:2010dp}, which may be elevated by 2-3 orders of magnitude
into adjacency with the Planck mass, where the $B_\mu = 0$ boundary condition fits like hand to glove~\cite{Ellis:2001kg,Ellis:2010jb,Li:2010ws}. This final $SU(5) \times U(1)_{\rm X}$ unification scale is denoted as $M_{\cal F}$, where we discover that for the viable parameter space consistent with the latest experiment, $M_{\cal F}$ emerges at about 4-6$\times 10^{17}$ GeV, right near the string scale of $\sim$5$\times 10^{17}$ GeV. This is in quite favorable contrast to the gap between the string scale and the traditional GUT scale of $\sim$2$\times 10^{16}$ GeV that introduced the ``little hierarchy'' problem. This natural resolution of the ``little hierarchy'' problem corresponds also to true string-scale gauge coupling unification in the free fermionic string models~\cite{Jiang:2006hf,Lopez:1992kg} or the decoupling scenario in F-theory models~\cite{Jiang:2009zza,Jiang:2009za}, and also helps to address the monopole problem via hybrid inflation.

The flippon multiplets trigger a similar effect on the RGEs of the gauginos as a result of the modifications to the $\beta$-function coefficients.  Particularly, the gaugino mass $M_{\rm 3}$ evolves down flat from the high energy boundary, obeying the relation $M_3/M_{1/2} \simeq \alpha_3(M_{\rm Z})/\alpha_3(M_{32}) \simeq \mathcal{O}\,(1)$, which elicits a noticeably light gluino mass.
In contrast, the $SU(2)_{\rm L}$ and hypercharge $U(1)_{\rm Y}$ associated gaugino masses are driven downward from the $M_{1/2}$ boundary value by approximately the ratio of their corresponding gauge couplings $(\alpha_2,\alpha_{\rm Y})$ to the strong coupling $\alpha_{\rm s}$.
A rather light stop squark $\widetilde{t}_1$ results from the large mass splitting expected from the heaviness of the top quark via its strong coupling to the Higgs (which is also key to generating an appreciable radiative Higgs mass shift $\Delta~m_h^2$~\cite{Okada:1990vk,Okada:1990gg,Haber:1990aw,Ellis:1990nz,Ellis:1991zd}). The characteristically predictive $m_{\tilde{t}_1} < m_{\tilde{g}} < m_{\tilde{q}}$ mass hierarchy of a light stop and gluino, both lighter than all the remaining squarks, is stable across the full No-Scale $\cal{F}$-$SU(5)$ model space, though not replicated in any CMSSM or mSUGRA constructions that we are aware of.

The spectrum associated with this mass hierarchy generates a unique event topology starting from the pair production of heavy squarks
$\widetilde{q} \widetilde{\overline{q}}$, except for the light stop, in the initial hard scattering process,
with each squark likely to yield a quark-gluino pair $\widetilde{q} \rightarrow q \widetilde{g}$.  Each gluino may be expected
to produce events with a high multiplicity of virtual stops or tops, via the (off-shell for $M_{1/2} < 729$ GeV) $\widetilde{g} \rightarrow \widetilde{t}_1 t$ transition, which in turn may terminate into hard scattering products such as $\rightarrow W^{+}W^{-} b \overline{b} \widetilde{\chi}_1^{0}$ and $W^{-} b \overline{b} \tau^{+} \nu_{\tau} \widetilde{\chi}_1^{0}$, where the $W$ bosons will produce mostly hadronic jets and some leptons. The model described may then consistently produce a net product of eight or more hard jets emergent from a single squark pair production event, passing through a single intermediate gluino pair, resulting after fragmentation in an impressive signal of ultra-high multiplicity final state jet events. 

The entirety of the viable $\cal{F}$-$SU(5)$ parameter space naturally features a dominantly bino LSP, at a purity greater than 99\%, as is exceedingly suitable for direct detection.  There exists no direct bino to wino mass mixing term. This distinctive and desirable model characteristic is guaranteed by the relative heaviness of the Higgs bilinear mass $\mu$, which in the present construction generically traces the the universal gaugino mass $M_{1/2}$ at the boundary scale $M_{\cal F}$, and subsequently transmutes under the RGEs to a somewhat larger value at the electroweak scale, as detailed in Table (\ref{tab:points}) for the ten representative points to be discussed later in this work.

A large region of the bare-minimally constrained~\cite{Li:2011xu} parameter space of No-Scale $\cal{F}$-$SU(5)$,
as defined by consistency with the world average top-quark mass $m_{\rm t}$, the No-Scale boundary conditions,
radiative EWSB, the centrally observed WMAP7 CDM relic density limits 0.1088 $\leq \Omega h^2 \leq$ 0.1158~\cite{Komatsu:2010fb} (we assume a thermal relic), and precision LEP constraints on the lightest CP-even Higgs boson $m_{h}$~\cite{Barate:2003sz,Yao:2006px} and other light SUSY chargino and neutralino mass content, remains viable even after careful comparison against the recent light Higgs boson mass constraints ~\cite{ATLAS:2012gk,CMS:2012gu,Aaltonen:2012qt}.  The intersection of these experimental bounds is quite non-trivial,
as the tight theoretical constraints, most notably the vanishing of $B_\mu$ at the high scale boundary, render the residual
parameterization insufficient for arbitrary tuning of even isolated predictions, not to mention the union of all predictions.

\begin{figure*}[htf]
        \centering
        \includegraphics[width=1.00\textwidth]{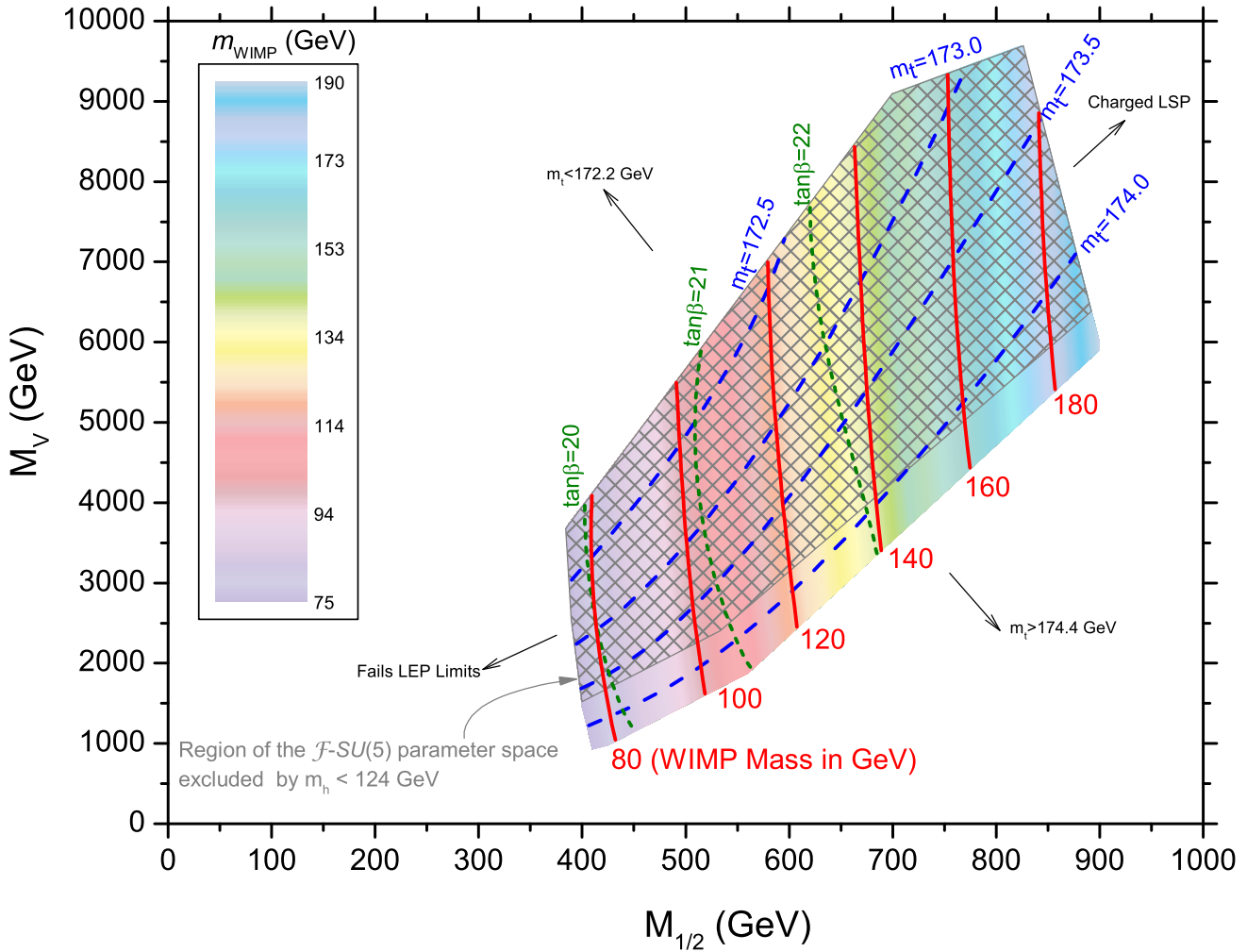}
        \caption{The bare-minimally constrained parameter space of No-Scale $\cal{F}$-$SU(5)$ is depicted as a function of the gaugino boundary mass $M_{1/2}$ and the vector-like mass $M_{\rm V}$. The WIMP mass, top quark mass $m_t$, and tan$\beta$ are demarcated via the solid, dashed, and dotted contour lines, respectively. The legend associates each shade with its respective numerical value of the WIMP mass, which is the lightest neutralino mass $m_{\widetilde{\chi}_1^0}$ in the present model. The four arrows positioned just outside the outer boundaries of the parameter space specify the reason for exclusion of the non-shaded regions. The region disfavored by an $m_h < 124$ GeV light Higgs boson is marked out with the crosshatch pattern. All masses are in GeV.}
        \label{fig:wimp_mass}
\end{figure*}

The cumulative result of the application of the bare-minimal constraints shapes the parameter space into the large profile situated in the $M_{1/2},M_{\rm V}$ plane exhibited in Fig.~(\ref{fig:wimp_mass}), from a more tapered light mass region with a lower bound of $\tan \beta$ = 19.4 demanded by the LEP constraints, into a more expansive heavier region that ceases sharply with the charged stau LSP exclusion around tan$\beta \simeq$ 23, where we overlay smooth contour gradients of top quark mass, tan$\beta$, and the WIMP mass. The bare-minimal constraints set lower bounds at about $M_{1/2} \simeq$ 385 and $M_V \simeq$ 925 GeV correlated to the lower bound on tan$\beta$ of around 19.4, and upper bounds near $M_{1/2} \simeq$ 900 and $M_V \simeq$ 9.7 TeV, correlated to the upper bound on tan$\beta$ at about 23. Though not visible in the two-dimensional rendering of Fig.~(\ref{fig:wimp_mass}), we note that the region in Fig.~(\ref{fig:wimp_mass}) representing the No-Scale $\cal{F}$-$SU(5)$ experimentally allowed model space consists of extended continuous regions of parameter space that could be regarded as layers, where the $B_\mu$=0 constraint at the high scale boundary and adherence to the 7-year WMAP limits on the relic density in our universe further constrain the region in Fig.~(\ref{fig:wimp_mass}) to that of a single layer. The model space beyond the hashed over region illustrated in Fig.~(\ref{fig:wimp_mass}) consists of those points within the parameter space not excluded by the recent light Higgs boson mass constraints~\cite{ATLAS:2012gk,CMS:2012gu,Aaltonen:2012qt}, as derived in Refs.~\cite{Li:2011xg,Li:2011ab,Li:2012jf} applying the additional contributions from radiative loops in the vector-like multiplets. The surviving region in Fig.~(\ref{fig:wimp_mass}) predicts a more narrowed region of the flippon vector-like mass 1 $\lesssim M_V \lesssim$ 6 TeV.

A top-down consistency condition on the gaugino boundary mass $M_{1/2}$, and the parametrically coupled value of $\tan\beta$,
is dynamically determined at a secondary local minimization $dV_{\rm min}/dM_{1/2}=0$ of the minimum of the Higgs potential $V_{\rm min}$.
Since $M_{1/2}$ is related to the modulus field of the internal string theoretic space, this also
represents a dynamic stabilization of the compactification modulus.  The result is demonstrably consistent with the bottom-up
phenomenological approach ~\cite{Li:2010uu,Li:2011dw,Li:2011xu,Li:2011ex}, and we note in particular a rather distinctive conclusion
that is enforced in both perspectives: the ratio $\tan\beta$ must have a value very close to 20.

A further noteworthy aspect emerging from our prior work on No-Scale $\cal{F}$-$SU(5)$ is a rather suggestive linkage between the Higgs bilinear mass term $\mu$ and the gaugino mass $M_{1/2}$. We find that $M_{1/2}$ is virtually equal to $\mu$ across the entire region of the model space investigated here, as evidenced in Table (\ref{tab:points}) for the ten representative points to be discussed shortly. This may be an effect of the strong No-Scale boundary conditions and might moreover have deep implications to the solution of the $\mu$ problem in the supersymmetric standard model~\cite{LMNW-P}. The fact that $\mu$ and $M_{\rm V}$ might be generated from the same mechanism~\cite{LMNW-P} represents an additional naturalness argument for the suggestion that $\mu$ and $M_{\rm V}$ should be of the same order.


\begin{figure*}[htf]
        \centering
        \includegraphics[width=0.65\textwidth]{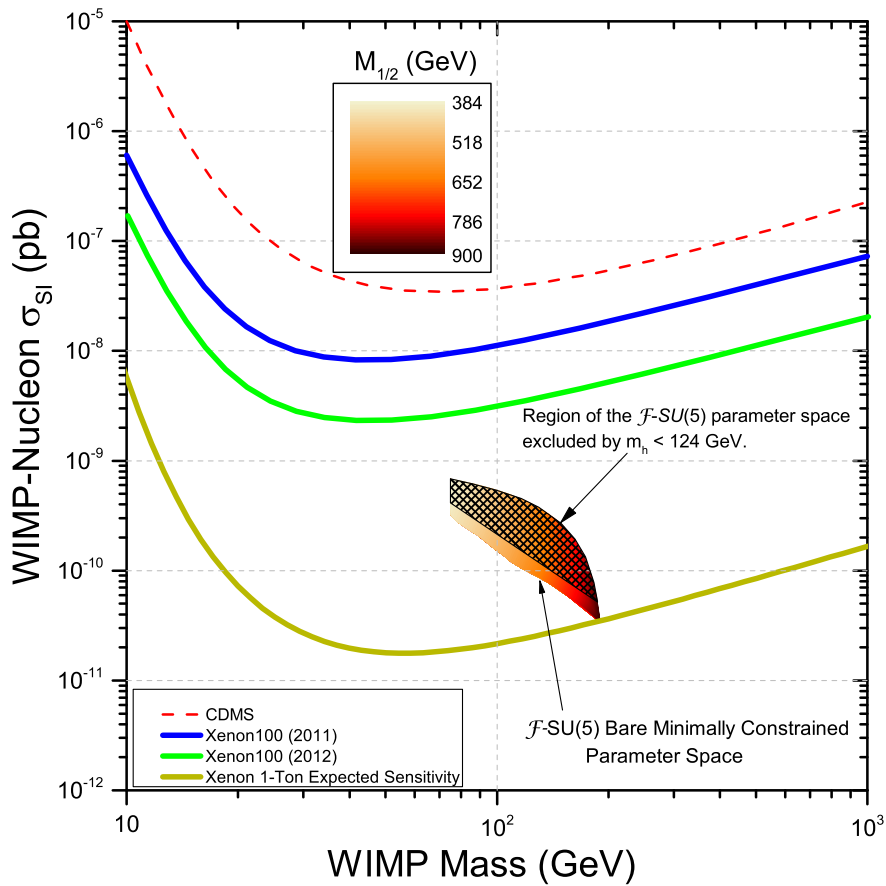}
        \caption{Direct dark matter detection diagram associating the WIMP mass with the spin-independent annihilation cross-section $\sigma_{\rm SI}$. Delineated are the current upper bounds from the CDMS~\cite{Ahmed:2009zw} and XENON100~\cite{Aprile:2011hi,Aprile:2012nq} experiments, including the projected sensitivity of the 1-ton XENON experiment~\cite{aprile}. The region of parameter space shown is that which satisfies the bare minimal phenomenological constraints of Ref.~\cite{Li:2011xu}, illustrated as a function of $M_{1/2}$ and $M_{V}$ in Fig.~(\ref{fig:wimp_mass}). The shading of the parameter space here depicts continuous levels of $M_{1/2}$. The region disfavored by an $m_h < 124$ GeV light Higgs boson is marked out with the crosshatch pattern. All masses are in GeV.}
        \label{fig:Xenon_sigma}
\end{figure*}

\begin{figure*}[htf]
        \centering
        \includegraphics[width=0.95\textwidth]{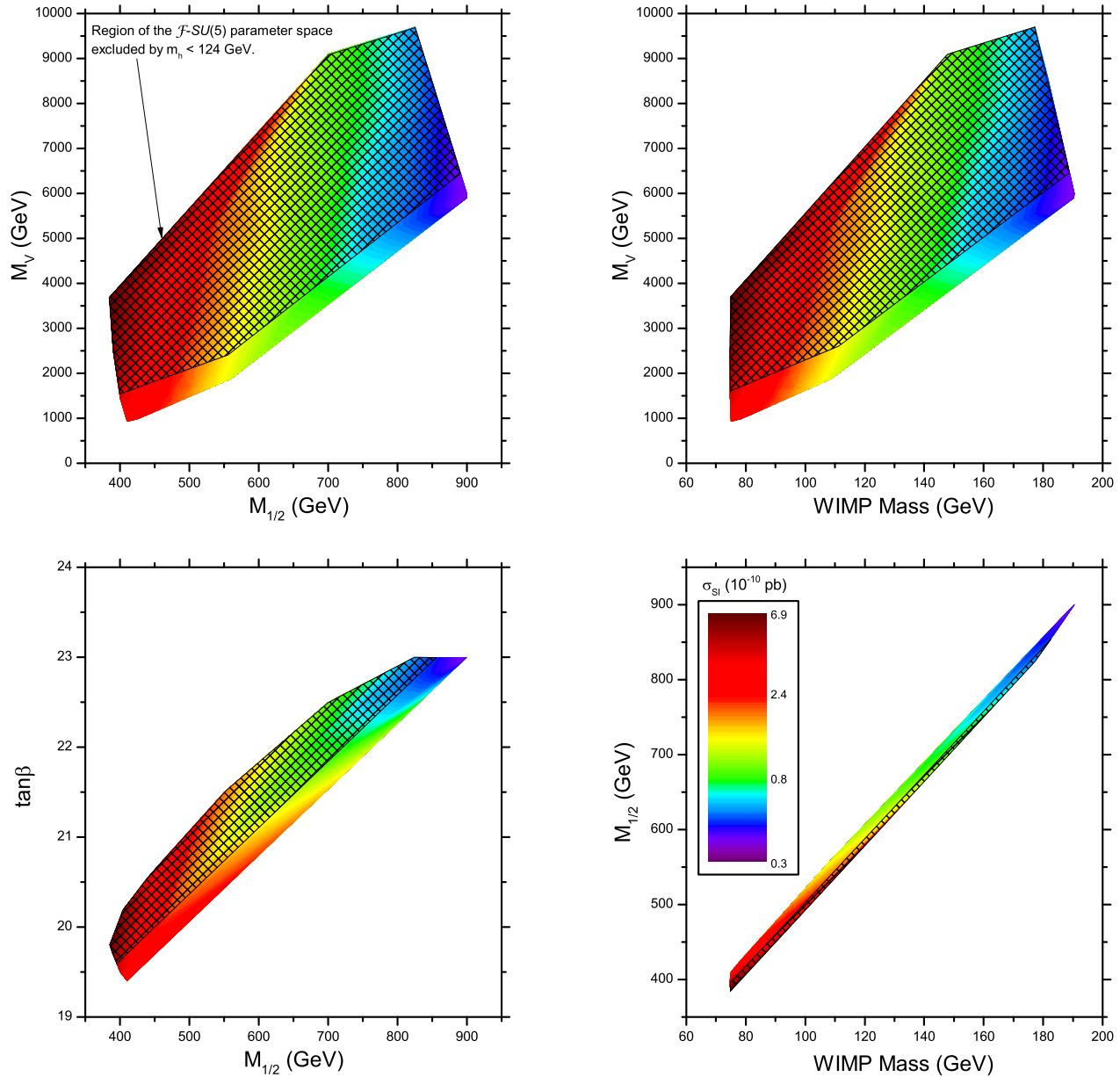}
        \caption{Shaded gradients linking the spin-independent annihilation cross-section $\sigma_{\rm SI}$ to the WIMP mass and model parameters, where the cross-section is essentially a flattened third dimension. The legend associates each shade with its respective numerical value of $\sigma_{\rm SI}$. The regions of parameter space depicted here are those which satisfy the bare minimal phenomenological constraints of Ref.~\cite{Li:2011xu}, illustrated as a function of $M_{1/2}$ and $M_{V}$ in Fig.~(\ref{fig:wimp_mass}). The region disfavored by an $m_h < 124$ GeV light Higgs boson is marked out with the crosshatch pattern. All masses are in GeV.}
        \label{fig:Xenon_4plex}
\end{figure*}

\begin{figure*}[htf]
        \centering
        \includegraphics[width=0.65\textwidth]{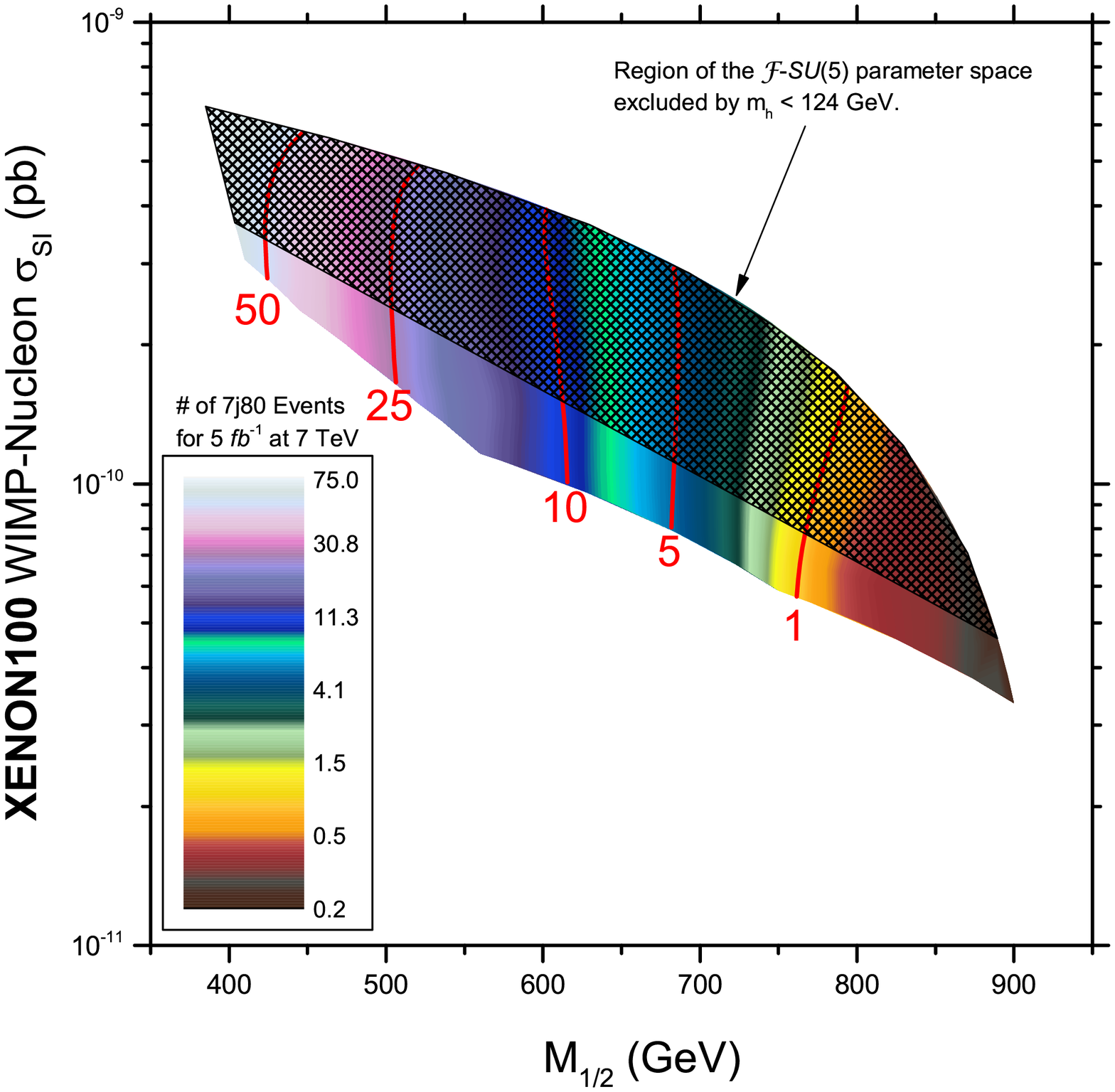}
        \caption{Correlation between the search for dark matter at the LHC with the XENON100 and all direct-detection experiments. The region exhibited here is the parameter space displayed in Figs.~(\ref{fig:wimp_mass})-(\ref{fig:Xenon_4plex}). The gradients represent the number of \fsu5 SUSY multijet events for $5~{\rm fb}^{-1}$ of luminosity at $\sqrt{s}$ = 7 TeV, using the ATLAS 7j80 search strategy of Ref.~\cite{ATLAS-CONF-2012-037}. The red contour lines demarcate the specifically labeled number of \fsu5 SUSY multijet events. The legend associates each shading with its respective numerical value of the number of \fsu5 SUSY multijet events. The region disfavored by an $m_h < 124$ GeV light Higgs boson is marked out with the crosshatch pattern. All masses are in GeV.}
        \label{fig:Xenon_LHC}
\end{figure*}


\section{A Dual Dark Matter Case Study}

The proportional rescaling associated with the single massive input $M_{1/2}$ explains the ability to generate the WMAP7 successfully and generically, where we assume a thermal relic.  The correct DM relic density can be generated by the LSP neutralino and light stau coannihilation. Moreover, because $M_{1/2}$=0 at tree-level, the gravitinos and moduli can be heavy in No-Scale supergravity, about $\sim$30 TeV~\cite{Linde:2011ja}. Hence, the decay of gravitinos and moduli into supersymmetry particles that could potentially repopulate the universe is negligible, and thus the reheating temperature of the early universe is small. All considered, it indicates how finely naturally adapted (not finely tuned) No-Scale $\cal{F}$-$SU(5)$ is with regards to the question of relic density. Although currently safe, it does appear that the full model space may be effectively probed in the near future by the extended reach of the ongoing data collection at XENON100 or one of its successors. The relevant scale dependent sensitivity contours to spin-independent DM-nucleon scattering, along with their relation to the putative $\cal{F}$-$SU(5)$ signal, are depicted in Fig.~(\ref{fig:Xenon_sigma}). The set of plots in Fig.~(\ref{fig:Xenon_4plex}) provide a detailed view of the interrelation between the gaugino mass $M_{1/2}$, the vector-like multiplet mass $M_V$, the ratio of Higgs vacuum expectation values (VEVs) $\tan \beta$, the bino WIMP mass, and the spin-independent DM-nucleon scattering cross section $\sigma_{\rm SI}$ in picobarns. Our SUSY mass spectrum and scattering cross-section calculations have been performed using {\tt MicrOMEGAs 2.4}~\cite{Belanger:2010gh}, employing a proprietary modification of the {\tt SuSpect 2.34}~\cite{Djouadi:2002ze} codebase to run the RGEs.

\begin{table*}[htbp]
	\centering
	\caption{Ten representative points selected from the viable parameter space depicted in Figs.~(\ref{fig:wimp_mass})-(\ref{fig:Xenon_LHC}), organized in terms of the parameters ($M_{1/2}$, $M_V$, tan$\beta$, $m_t$). The WIMP mass is represented by $m_{\widetilde{\chi}_1^0}$, which is the lightest neutralino mass in the present model. The values of the $\mu$ parameter are those at the electroweak symmetry breaking scale and at the unification scale $M_{\cal F}$. The relic density $\Omega h^2$ constraint strictly adheres to the 7-year WMAP measurements 0.1088 $\leq \Omega h^2 \leq$ 0.1158~\cite{Komatsu:2010fb}. The $\sigma_{\rm SI}$ values represent the spin-independent annihilation cross-sections. The final column specifies the number of \fsu5 SUSY multijet events for 5 \fb at $\sqrt{s}$ = 7 TeV, using the ATLAS 7j80 search strategy of Ref.~\cite{ATLAS-CONF-2012-037} to exhibit the close correlation in No-Scale \fsu5 between multijet events at the LHC and the spin-independent cross-section. All masses specified here are in GeV.}
		\begin{tabular}{c|c|c|c||c|c|c|c|c|c} \hline
$~~M_{1/2}~~$&$~~~M_{\rm V}~~~$&$~~\tan\beta~~$&$~~~m_{t}~~~$&$~~m_{\chi^{0}_{1}}~~$&$~\mu({\rm EW})~$&$~\mu(M_{\cal F})~$&$~~~\Omega h^2~~~$&$~\sigma_{\rm SI}~(\times10^{-10}{\rm pb})~$&${\cal F}-SU(5)~{\rm 7j80~Events}$ \\ \hline \hline	
$	405	$&$	1050	$&$	19.40	$&$	173.8	$&$	77	$&$	725	$&$	409	$&$	0.1126	$&$	3.8	$&$	51.0	$	\\	\hline
$	450	$&$	1250	$&$	19.95	$&$	174.0	$&$	87	$&$	788	$&$	453	$&$	0.1120	$&$	2.6	$&$	41.4	$	\\	\hline
$	505	$&$	1370	$&$	20.47	$&$	174.2	$&$	98	$&$	868	$&$	506	$&$	0.1131	$&$	1.7	$&$	27.1	$	\\	\hline
$	550	$&$	1800	$&$	20.85	$&$	174.3	$&$	109	$&$	924	$&$	551	$&$	0.1113	$&$	1.3	$&$	19.1	$	\\	\hline
$	610	$&$	1985	$&$	21.25	$&$	174.4	$&$	123	$&$	1007	$&$	607	$&$	0.1134	$&$	0.95	$&$	10.6	$	\\	\hline
$	650	$&$	2225	$&$	21.48	$&$	174.4	$&$	132	$&$	1059	$&$	645	$&$	0.1139	$&$	0.78	$&$	6.9	$	\\	\hline
$	708	$&$	2612	$&$	21.83	$&$	174.4	$&$	145	$&$	1132	$&$	699	$&$	0.1110	$&$	0.61	$&$	4.1	$	\\	\hline
$	780	$&$	3886	$&$	22.38	$&$	174.4	$&$	164	$&$	1209	$&$	766	$&$	0.1132	$&$	0.49	$&$	1.0	$	\\	\hline
$	850	$&$	4310	$&$	22.52	$&$	174.4	$&$	181	$&$	1300	$&$	832	$&$	0.1129	$&$	0.38	$&$	0.4	$	\\	\hline
$	900	$&$	5000	$&$	22.75	$&$	174.4	$&$	193	$&$	1357	$&$	877	$&$	0.1094	$&$	0.33	$&$	0.2	$	\\	\hline
		\end{tabular}
		\label{tab:points}
\end{table*}

\begin{table*}[htbp]
	\centering
	\caption{Supersymmetry and light Higgs boson masses of the ten representative points of Table~(\ref{tab:points}), organized in terms of the parameters ($M_{1/2}$, $M_V$, tan$\beta$, $m_t$). The second lightest neutralino $m_{\widetilde{\chi}_{2}^{0}}$ and lightest chargino $m_{\widetilde{\chi}_{1}^{\pm}}$ masses are degenerate in the present model, which we designate here in the short-hand notation $m_{\chi^{0}_{2}/\chi^{\pm}_{1}}$. Note the mass splitting between the gluino $\widetilde{g}$ and light stop $\widetilde{t}_1$, which is natural in No-Scale $\cal{F}$-$SU(5)$, the transition of which requires a virtual stop for $M_{1/2} < 729$ GeV, via $\widetilde{g} \rightarrow  \widetilde{t}_1 t$. All masses specified here are in GeV.}
		\begin{tabular}{c|c|c|c||c|c|c|c|c|c|c|c|c|c|c|c|c} \hline
$M_{1/2}$&$~M_{\rm V}~$&$~\tan\beta~$&$~m_{t}~$&$~m_{\chi^{0}_{2}/\chi^{\pm}_{1}}~$&$~m_{\widetilde{\tau}_{1}}~$&$~m_{\widetilde{e}_{R}}~$&$~m_{\widetilde{t}_{1}}~$&$~m_{\widetilde{t}_{2}}~$&$~m_{\widetilde{b}_{1}}~$&$~m_{\widetilde{b}_{2}}~$&$~m_{\widetilde{u}_{R}}~$&$~m_{\widetilde{u}_{L}}~$&$~m_{\widetilde{d}_{R}}~$&$~m_{\widetilde{d}_{L}}~$&$~~m_{\widetilde{g}}~~$&$~~m_h~~$ \\ \hline \hline	
$	405	$&$	1050	$&$	19.40	$&$	173.8	$&$	167	$&$	86	$&$	155	$&$	415	$&$	802	$&$	742	$&$	840	$&$	841	$&$	914	$&$	876	$&$	918	$&$	552	$&$	125.7	$	\\	\hline
$	450	$&$	1250	$&$	19.95	$&$	174.0	$&$	188	$&$	96	$&$	171	$&$	471	$&$	870	$&$	815	$&$	917	$&$	921	$&$	1002	$&$	959	$&$	1005	$&$	611	$&$	125.4	$	\\	\hline
$	505	$&$	1370	$&$	20.47	$&$	174.2	$&$	213	$&$	107	$&$	191	$&$	538	$&$	955	$&$	906	$&$	1015	$&$	1022	$&$	1111	$&$	1063	$&$	1114	$&$	681	$&$	126.1	$	\\	\hline
$	550	$&$	1800	$&$	20.85	$&$	174.3	$&$	235	$&$	118	$&$	207	$&$	591	$&$	1019	$&$	974	$&$	1085	$&$	1095	$&$	1192	$&$	1138	$&$	1194	$&$	742	$&$	125.0	$	\\	\hline
$	610	$&$	1985	$&$	21.25	$&$	174.4	$&$	263	$&$	131	$&$	228	$&$	663	$&$	1112	$&$	1071	$&$	1188	$&$	1201	$&$	1308	$&$	1248	$&$	1310	$&$	818	$&$	125.5	$	\\	\hline
$	650	$&$	2225	$&$	21.48	$&$	174.4	$&$	283	$&$	140	$&$	243	$&$	709	$&$	1172	$&$	1134	$&$	1254	$&$	1270	$&$	1383	$&$	1319	$&$	1385	$&$	871	$&$	125.4	$	\\	\hline
$	708	$&$	2612	$&$	21.83	$&$	174.4	$&$	311	$&$	153	$&$	264	$&$	777	$&$	1259	$&$	1224	$&$	1348	$&$	1367	$&$	1489	$&$	1420	$&$	1491	$&$	945	$&$	125.3	$	\\	\hline
$	780	$&$	3886	$&$	22.38	$&$	174.4	$&$	348	$&$	171	$&$	290	$&$	859	$&$	1357	$&$	1326	$&$	1452	$&$	1476	$&$	1608	$&$	1531	$&$	1610	$&$	1044	$&$	124.3	$	\\	\hline
$	850	$&$	4310	$&$	22.52	$&$	174.4	$&$	382	$&$	188	$&$	315	$&$	937	$&$	1464	$&$	1436	$&$	1568	$&$	1596	$&$	1739	$&$	1655	$&$	1740	$&$	1133	$&$	124.7	$	\\	\hline
$	900	$&$	5000	$&$	22.75	$&$	174.4	$&$	408	$&$	200	$&$	333	$&$	993	$&$	1536	$&$	1510	$&$	1645	$&$	1675	$&$	1825	$&$	1736	$&$	1827	$&$	1199	$&$	124.7	$	\\	\hline
		\end{tabular}
		\label{tab:masses}
\end{table*}

We take a sufficiently broad sampling of the viable No-Scale $\cal{F}$-$SU(5)$ parameter space such that a global picture of the model's phenomenology can emerge. Fig.~(\ref{fig:Xenon_LHC}) demonstrates that a correlation does exist in this model between the collider and direct detection DM search methodologies. To clearly exhibit this correlation, we directly compare the number of \fsu5 SUSY multijet events for $5~{\rm fb}^{-1}$ at the $\sqrt{s}$= 7 TeV LHC for the ATLAS search strategy of Ref.~\cite{ATLAS-CONF-2012-037} against the gaugino mass $M_{1/2}$ and also the spin-independent DM-nucleon scattering cross section $\sigma_{\rm SI}$ in picobarns. As Fig.~(\ref{fig:Xenon_LHC}) illustrates, there is a prominent correlation between the outlook of these two key experiments, with an enhancement in the prospect for near-term DM detection occurring, as might be expected, at the lighter mass scales in both cases. For our collider level analysis, we enumerate in Fig.~(\ref{fig:Xenon_LHC}) the number of \fsu5 SUSY events for the $5~{\rm fb}^{-1}$ delivered at $\sqrt{s} = 7$ TeV. We mimic the ATLAS post-processing cuts of Ref.~\cite{ATLAS-CONF-2012-037} for the case of jet $p_T >$ 80 GeV and $H_{\rm T}^{\rm miss}/\sqrt{H_{\rm T}}$ in the range $(4.0 \to \inf)$. We retain only those events with greater than or equal to seven jets, where we refer to this search strategy as 7j80, due to the more suppressive nature of the ATLAS cuts in the high jet multiplicity regime.

We execute on each of the ten benchmark samples an in-depth Monte Carlo collider-detector simulation of all 2-body SUSY
processes based on the {\tt MadGraph}~\cite{Stelzer:1994ta,MGME} program suite, including the {\tt MadEvent}~\cite{Alwall:2007st},
{\tt PYTHIA}~\cite{Sjostrand:2006za} and {\tt PGS4}~\cite{PGS4} chain. The SUSY particle masses and spin-independent cross-sections are calculated with {\tt MicrOMEGAs 2.4}, applying the proprietary modification of the {\tt SuSpect 2.34} codebase to run the flippon-enhanced RGEs. We implement a modified version of the default ATLAS detector specification card provided with {\tt PGS4} that
calls an anti-kt jet clustering algorithm, indicating an angular scale parameter of $\Delta R = 0.4$.  The resultant event files are filtered according to a precise replication of the selection cuts specified by the ATLAS Collaboration in Ref.~\cite{ATLAS-CONF-2012-037}, employing the script {\tt CutLHCO 2.0}~\cite{Walker:2012vf} to implement the post-processing cuts. 

We further observed in the simulation a strongly peaked Poisson-like distribution of the histogram on missing transverse energy, defined as:
\begin{equation}
H_{\rm T}^{\rm miss} \equiv
\sqrt{{\left( \sum_{\rm jets} p_{\rm T} \cos \phi \right)}^2 + {\left( \sum_{\rm jets} p_{\rm T} \sin \phi \right)}^2}\quad.
\label{HTmiss}
\end{equation}
Curiously, the central value appears to correlate strongly with a small multiple of the LSP neutralino mass. Given the allowed kinetic excess over the jet invariant mass, the trigonometric reduction from extraction of the beam-transverse component, and the possibility of partial cancellation between the dual neutralino signal (with randomly oriented directionality due to the multi level decay cascade), it does not seem obvious that this should be the case. We have devoted a full investigation to the theoretical origins and broader generality and persistence of this result across the parameter
space and with various beam energies in a complementary publication~\cite{Li:2011gh}, including also comparisons against CMSSM model control samples. Indications suggest a possibly effective tool for extraction of the LSP mass from collider level events, allowing for a richer empirical correlation against direct detection observations. We emphasize that the suppression of the SM background necessary for the potential extraction of such a peak is naturally provided by a multijet search methodology.


We exact out of the viable region of parameter space displayed in Figs.~(\ref{fig:wimp_mass})-(\ref{fig:Xenon_LHC})
ten representative points in Table (\ref{tab:points}) and Table (\ref{tab:masses}), where we identify the points by their respective $M_{1/2}$, $M_{\rm V}$, $\tan \beta$, and $m_{t}$. In Table (\ref{tab:points}) we provide the WIMP mass, the $\mu$ parameter at the electroweak scale and unification scale $M_{\cal F}$, relic density $\Omega h^2$, spin-independent annihilation cross-section $\sigma_{\rm SI}$, and the total number of \fsu5 SUSY 7j80 multijet events for $5~{\rm fb}^{-1}$ of luminosity at 7 TeV. Details of the SUSY spectrum for these ten points are shown in Table (\ref{tab:masses}). Notice the mass splitting between the gluino mass $m_{\widetilde{g}}$ and light stop mass $m_{\widetilde{t}_1}$ in Table (\ref{tab:masses}), the transition of which requires a virtual light stop, proceeding through the decay (off shell for $M_{1/2} < 729$ GeV) process $\widetilde{g} \rightarrow  \widetilde{t}_1 t$, where the light stops lead to additional top quarks. Thus, the source of the large number of multijet events via the production of numerous top quarks occurs very elegantly and naturally in No-Scale $\cal{F}$-$SU(5)$, concurrently with the naturally generated 125 GeV lightest Higgs boson mass.

The discrete numerical data in Table (\ref{tab:points}) and continuous gradient contours
in Fig.(\ref{fig:Xenon_LHC}) provide compelling linkage between the XENON100 experiment and the LHC in
No-Scale $\cal{F}$-$SU(5)$.  The bino mass range for our parameter space is from about 75 to 200 GeV,
where most of this LSP range remains beyond the boundaries explored thus far at the colliders and direct detection experiments.
The predominantly bino composition of the LSP follows from the large Higgs mixing mass $\mu$ at the electroweak
scale of about 700 GeV to 1.4 TeV, as specified in Table (\ref{tab:points}).


\section{Conclusions}

The immediate era of particle physics research represents a historic opportunity for the experimentalist, the phenomenologist, and the theorist, as the decades long pursuit of a discovery and correct description of dark matter potentially advances toward an ultimate resolution in the near future.  On the experimental front, there are two key approaches to the SUSY DM search, with the collider production efforts led currently by the LHC, and the best near term hopes of direct detection carried by the XENON100 collaboration. Though the two approaches are quite distinct, their fundamental goals are identical and their empirical scopes are complementary.

The search for a dark matter candidate is also transpiring on the model building front for a picture which may efficiently and naturally satisfy all current experimental bounds, while making clearly testable predictions, ideally within the scope of the near term experiments.  We have continued our description of the model named No-Scale $\cal{F}$-$SU(5)$, which we claim actualizes each of these necessary and desirable characteristics.  We have undertaken a dual case study of the specific signatures which a No-Scale $\cal{F}$-$SU(5)$ SUSY signal would demonstrate at both XENON100 and the LHC. 

In our model, the LSP neutralino is over 99\% bino due to the relatively large $\mu$ term at the electroweak scale.  Thus, the viable parameter space automatically satisfies the current constraints from the XENON100 and CDMS/EDELWEISS experiments. However, the signal is not dramatically beneath the current detection threshold of the direct detection experiments, so that we consider future detectability to be quite plausible. Moreover, the viable parameter space can be tested at the early LHC run by searching for a supersymmetry signal of ultra-high multiplicity hadronic jets which are a characteristic feature of the distinctive No-Scale $\cal{F}$-$SU(5)$ mass hierarchy.  We add that the predicted vector-like particles may also themselves be directly produced during the ongoing and future LHC runs. The lightest CP-even Higgs boson mass in the Minimal Supersymmetric Standard Model is consistently predicted to be about 120 GeV, and with the additional contributions from the vector-like particles, a 125 GeV higgs boson mass can be naturally generated.

The ability of No-Scale $\cal{F}$-$SU(5)$ to satisfy current experiment and remain largely intact from advancing LHC constraints is made all the more persuasive by comparison to the standard mSUGRA based alternatives, which despite a significantly greater freedom of parameterization, was rapidly cut down by the early results from the LHC and from XENON100.


\begin{acknowledgments}
This research was supported in part 
by the DOE grant DE-FG03-95-Er-40917 (TL and DVN),
by the Natural Science Foundation of China 
under grant numbers 10821504 and 11075194 (TL),
by the Mitchell-Heep Chair in High Energy Physics (JAM),
and by the Sam Houston State University
2011 Enhancement Research Grant program (JWW).
We also thank Sam Houston State University
for providing high performance computing resources.
\end{acknowledgments}


\bibliography{bibliography}

\end{document}